\documentclass[12pt]{JHEP3}

\usepackage{cite}
\usepackage{dsfont}
\usepackage{amsfonts}
\usepackage{amssymb,latexsym}
\usepackage{amsmath,amsbsy}
\usepackage{epsfig,bm}
\usepackage{graphicx,comment}
\usepackage{yfonts}

\def\cA{{\cal A}}  \def\cC{{\cal C}} 
   
   \def\cL{{\cal L}}
\def\cM{{\cal M}}  \def\cO{{\cal O}}

\def\Im{\mathop{\rm Im}}
\def\Re{\mathop{\rm Re}}

\newcommand{\tmfloatcontents}{}
\newlength{\tmfloatwidth}
\newcommand{\tmfloat}[5]{
  \renewcommand{\tmfloatcontents}{#4}
  \setlength{\tmfloatwidth}{\widthof{\tmfloatcontents}+1in}
  \ifthenelse{\equal{#2}{small}}
    {\ifthenelse{\lengthtest{\tmfloatwidth > \linewidth}}
      {\setlength{\tmfloatwidth}{\linewidth}}{}}
    {\setlength{\tmfloatwidth}{\linewidth}}  \begin{minipage}[#1]{\tmfloatwidth}
    \begin{center}
      \tmfloatcontents
      \captionof{#3}{#5}
    \end{center}
  \end{minipage}}

\newcommand{\sh}[1]{#1\hskip-7pt \diagup}
\newcommand{\e}{\epsilon}

\newcommand{\beq}{\begin{equation}}
\newcommand{\eeq}[1]{\label{#1}\end{equation}}
\newcommand{\bea}{\begin{eqnarray}}
\newcommand{\eea}[1]{\label{#1}\end{eqnarray}}

\title{Analyticity, Unitarity and One-loop Graviton Corrections to Compton Scattering}

\author{Hovhannes~R.~Grigoryan \\
\vspace{0.1in}

Center for Cosmology and Particle Physics\\
Department of Physics, New York University\\
4 Washington Place, New York, NY 10003, USA \\~~\\

E-mail addresses: \email{hg41@nyu.edu}

\vspace{0.1in}
}

\abstract{ 
We compute spin-flip cross section for graviton photoproduction on a spin-1/2 target of finite mass.
Using this tree-level result, we find one-loop graviton correction to the spin-flip low-energy forward Compton scattering amplitude by using Gerasimov-Drell-Hearn sum rule. We show that this result agrees with the corresponding perturbative computations, implying the validity of the sum rule at one-loop level, contrary to the previous claims.
We discuss possible effects from the black hole production and string Regge trajectory exchange at very high energies. These effects seem to soften the UV divergence present at one-loop graviton level. 
Finally, we discuss the relation of these observations with the models that involve extra dimensions.
}

\keywords{Anomalous magnetic moment, Compton scattering, gravity, sum rule, unitarity}

\begin{document}

\section{Introduction}

It has been known for a long time that in computing the one-loop graviton corrections to the anomalous magnetic moment of a charged lepton miraculous cancellations occur, rendering to a finite result. In four dimensions this has been shown first by Berends and Gastmans \cite{Berends:1974tr}, and farther generalized to certain extra dimensional model by Graesser \cite{Graesser:1999yg}. 
On the other hand, it is well known that gravity coupled to QED can be considered at most as an effective field theory. 
At the same time, analyticity and unitarity along with certain assumptions about the asymptotic behavior of the scattering amplitudes may provide useful constraints on the low energy coefficients of this effective theory (in the spirit of Ref.~\cite{Adams&al}). One of such constraints follows from the well known Gerasimov-Drell-Hearn (GDH) sum rule \cite{Gerasimov:1965et} that relates the anomalous magnetic moment to a certain dispersion integral.
Surprisingly, it has been argued in \cite{Goldberg:1999gc} that the GDH sum rule does not hold at the one-loop graviton level, either in four or in any number of extra dimensions. This posed a serious problem that was left without attention for more than a decade.

Here we show that the GDH sum rule is in fact satisfied at the one-loop graviton level. This requires the incorporation of certain gravity specific interactions that were ignored before. The resolution of this problem shows that the low energy effective theory of gravity coupled to QED at one-loop level is not in any obvious contradiction with either analyticity or unitarity, as was previously thought. However, clearly, this does not mean that analyticity or unitarity are maintained at multiple loop level or far from the perturbative regime.


\textit{Generalities:} Given a particle of mass $m$ and electric charge $e$, the Gerasimov-Drell-Hearn sum rule \cite{Gerasimov:1965et} connects the gyromagnetic ratio $g$ to a dispersion integral.\footnote{As usual, $g$ is defined as the ratio of the particle's magnetic moment $\mu$ to its spin $J$, so that:
$\vec{\mu} = \frac{e g}{2m}\vec{J}$.}
The main ingredient of the sum rule (see e.g. \cite{Low:1954kd,Weinberg}) is the low energy 
forward\footnote{The photon propagates along some $z$-direction with helicity $\lambda = \pm 1$,  and the target has a spin-$z$ projection $J_z$.} 
Compton scattering amplitude of a photon with
energy $\omega$ and helicity $\lambda$ off a massive target of spin $J$. This amplitude, $f_{\rm scat} (\omega, \lambda)$, is a real analytic function of the  photon's energy $\omega$ away from 
the real $\omega$-axis, where cuts and poles may exist at $\omega > 0 $. The imaginary part of $f_{\rm scat}$  is given 
by the optical theorem:
\begin{align}\label{opttheorem}
{\rm Im} f_{\rm scat} (\omega,\lambda) = \frac{\omega}{4\pi}\sigma_{\rm tot}(\omega, \lambda) \ ,
\end{align}
where $\sigma_{\rm tot}$ is the total cross-section for a photon with helicity $\lambda$ and energy $\omega$. 
Define now the following function:
\begin{align}\label{minusamplitude}
&f_-(\omega^2) \equiv\frac{f_{\rm scat} (\omega,+1)-f_{\rm scat} (\omega,-1)}{2\omega} \ .
\end{align}
As was discussed in Ref.~\cite{Grigoryan:2012xv}, when no intermediate (one particle) state exists in the Compton scattering, with either mass or spin different from those of the target then, it can be checked that:
\begin{align}\label{GDH0}
&f_-(\omega^2 \to 0)  = \frac{\alpha}{m^2}J_z a^2  \ .
\end{align}
where $a \equiv (g-2)/2$ is the anomalous magnetic moment (as usual $\alpha = e^2/4\pi$).
Using the optical theorem (\ref{opttheorem}) and definition (\ref{minusamplitude}), we have:
\begin{align}\label{opt2}
{\rm Im}f_-(\omega^2) = \frac{1}{8\pi}\Delta\sigma (\omega) \ , \qquad
\Delta\sigma (\omega) \equiv \sigma_{\rm tot}(\omega,+1) - \sigma_{\rm tot}(\omega,-1)   \ .
\end{align}
Assuming that $f_-(\omega^2)$ vanishes when $|\omega^2| \to \infty$, one can write an {\em unsubtracted} dispersion relation:
\begin{align}\label{dispGDH}
f_-(\omega^2) = \frac{1}{4\pi^2}\int^{\infty}_0\frac{\Delta\sigma (\omega')}{\omega'^2 - \omega^2 - i \epsilon} ~\omega' d\omega' \ .
\end{align}
When $\omega^2=0$, and $J_z=1/2$, Eqs.(\ref{GDH0}) and (\ref{dispGDH}), imply the 
GDH sum rule \cite{Gerasimov:1965et}:
\begin{align}\label{GDH}
a^2 = \frac{m^2}{2\pi^2 \alpha }\int^{\infty}_0\frac{\Delta\sigma(\omega')}{\omega'} d\omega'  \ .
\end{align}
In weakly interacting systems the RHS of Eq.~(\ref{GDH}) is expected to be $O(\alpha)$, since the lowest order contribution to the cross section may seem to come from the process $\gamma \ell \to \gamma \ell$,\footnote{Here by $\ell$ we mean some charged lepton (as usual $\gamma$ represents a photon).} thus, $\Delta \sigma_{\gamma \ell \to \gamma \ell} \sim \cO(\alpha^2)$. However, from the perturbative QED computations it is known that the LHS is of order $\cO(\alpha^2)$. This suggest that the RHS should vanish for the process $\gamma \ell \to \gamma \ell$, considered at the tree-level, and the leading contribution to the integral should come from the processes for which $\Delta \sigma \sim \cO(\alpha^3)$. In what follows, it will be instructive to show how this happens in more details.

The asymmetry in the differential Compton cross section is:\footnote{See, e.g., L.~D.~Landau, E.~M.~Lifshitz, ``Quantum Electrodynamics,'' Vol.IV, Ch.10.}
\begin{align}\label{diffasymCompton}
\left(\frac{d\sigma}{d\Omega}\right)_{\lambda=1} -  \ \left(\frac{d\sigma}{d\Omega}\right)_{\lambda=-1} &= -\frac{\alpha^2}{m^2}\left(\frac{\omega'}{\omega}\right)^2\frac{\omega + \omega'}{m}\cos \theta (1-\cos\theta) \ ,  \\[7pt] \nonumber
\omega' &= \frac{\omega}{1 + \frac{\omega}{m}(1-\cos \theta)} \ ,
\end{align}
where $\theta$ is a scattering angle in the laboratory frame.
Integrating over $d\Omega = 2\pi \sin\theta d \theta$, we will obtain, the total asymmetry in the cross section to be:
\begin{align}
\Delta \sigma_{\rm Compton}(\omega) = -\frac{2\pi \alpha^2}{m \omega}\left[\left(1+ \frac{m}{\omega}\right)\ln\left(1+\frac{2\omega}{m}\right) - 2\left(1+\frac{\omega^2}{(m+2\omega)^2} \right) \right] \ .
\end{align}
Substituting this expression inside the GDH integral (\ref{GDH}), one would indeed obtain zero! This non trivial observation was made in Ref.~\cite{Altarelli:1972nc}. Much later, it has been shown in Ref.~\cite{Dicus:2000cd} that the relevant order $\cO(\alpha^3)$ corrections to $\Delta\sigma$ are coming from the following two processes: a) $\gamma \ell \to \gamma \ell $, at order $\cO(\alpha^3)$, that is from the Compton scattering at one-loop \cite{Brown:1952eu}, interfering with the tree-level process, and b) from the process $\gamma\ell \to \gamma\gamma \ell$. Moreover, the computation of the total asymmetry in Ref.~\cite{Dicus:2000cd}  reproduced the correct result for $a$, known from the perturbation theory.

The paper is organized as follows:
in Sec.II, we demonstrate the reasoning that led to an incorrect conclusion that GDH sum rule does not hold  at the one-loop graviton level. For this we compute the asymmetry in the polarized graviton photoproduction cross section on a target of spin-1/2 and finite mass. Substituting this result in the GDH integral, and applying the known result of Berends and Gastmans, we arrive at a contradiction.
In Sec.III, we demonstrate that the problem is resolved if one incorporates gravity specific one-loop diagrams that contribute accordingly to the spin-flip forward Compton scattering amplitude.
In Sec.IV, we carry semi-qualitative discussions on possible effects from the nonperturbative regime, where either the exchange of the black hole or string states is important.
In Sec.V, we extend the arguments for models with large extra dimensions, and show that in 4D the collinear singularity and the UV divergence are related.
Finally, we summarize the obtained results.

\section{The Origin of the Problem}

\subsection{One-loop Graviton Contribution}

In Ref.~\cite{Berends:1974tr}, Berends and Gastmans found that the total one-loop quantum gravity correction to the anomalous magnetic moment $a = (g-2)/2$ of charged leptons give the following finite result:\footnote{Only QED and quantum gravity sectors were considered. Inclusion of the whole Standard Model is also possible.}
\begin{align}\label{BG}
a^{QG} = \frac{7}{4 \pi}G_N m^2 =  \frac{7}{16\pi^2}\frac{m^2}{M^2_P}  \ ,
\end{align}
where $m$ is the mass of the lepton, $G_N = 1/(4\pi M^2_P)$ is the 4D Newton's constant, and $M_P$ is Planck's constant. In particular, for muon $a^{QG}_{\mu}  \approx 4.2 \times 10^{-41}$. The relevant diagrams are shown in Fig.~\ref{QGdiagr}. Clearly, even including all other contributions from the Standard Model (SM), this result does not have any chance to be tested experimentally.\footnote{The anomalous magnetic moment of the muon is well measured as is \cite{PDG12}:
%
$a_{\mu}^{\rm exp} = 11 659 208.9(5.4)(3.3) \times 10^{-10}$,
%
in remarkable agreement with the SM prediction: $a_{\mu}^{\rm SM}  = 11659180.2(2)(42)(26) \times 10^{-10}$.}
%


\FIGURE{\includegraphics[width=12cm]{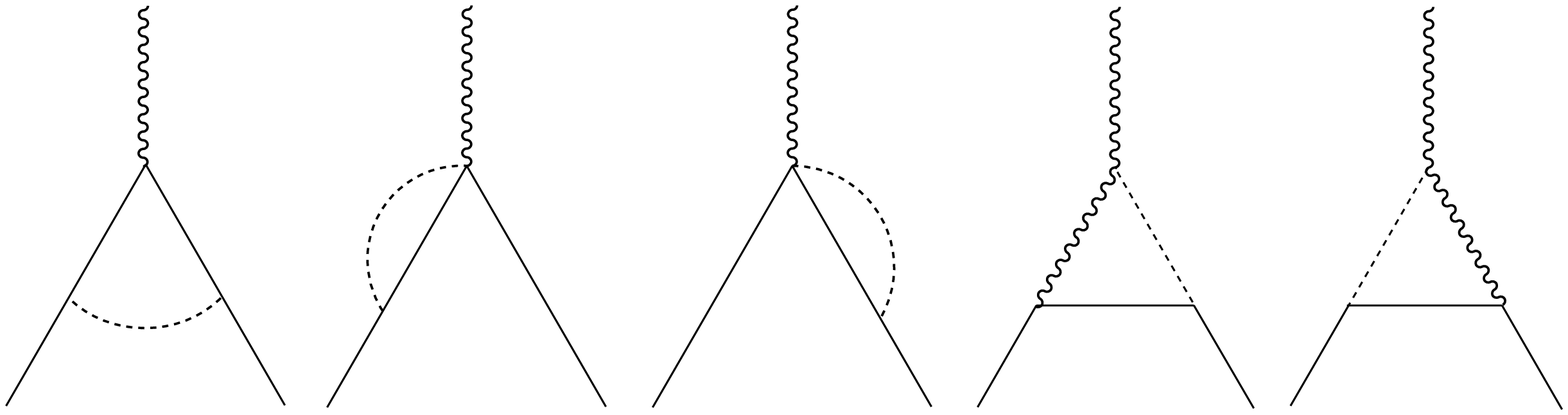}
\caption{One-loop graviton diagrams that contribute to the anomalous magnetic moment of lepton. The wavy lines represent photons, dashed lines - gravitons, and solid lines - leptons.}
\label{QGdiagr}
}

However, as was noticed by Goldberg \cite{Goldberg:1999gc}, this result (\ref{BG}) has an interesting theoretical significance. It appears that naive application of the GDH sum rule, when gravity sector is involved, may lead to problems.
For simplicity, we consider only the QED sector of the SM, and gravity. It may seem natural to expect that at the one-loop level the corrections to $a$ from these sectors should be additive: $a = a^{QED} + a^{QG}$. As a result, at the one-loop level, the GDH sum rule can be written as:\footnote{See also Ref.~\cite{Brodsky:1968ea} where similar ideas were applies if instead of QG some other new physics is involved.}
\begin{align}
(a^{QED})^2 + 2a^{QED} a^{QG} + (a^{QG})^2 = \frac{m^2}{2\pi^2 \alpha}\int^{\infty}_{0} \frac{d\omega}{\omega}\left[\Delta\sigma^{QED}(\omega) + \Delta\sigma'(\omega) \right] \ ,
\end{align}
where $a^{QED} \approx \alpha/2\pi$, the first term in the integral, $\Delta\sigma^{QED}(\omega)$, is purely QED contribution while the second, $\Delta\sigma'(\omega)$, is the contribution from both gravity and QED sectors. Since QED by itself satisfies the sum rule, we are left with:
\begin{align}\label{GDHint}
2a^{QED} a^{QG} + (a^{QG})^2 = \frac{m^2}{2\pi^2 \alpha}\int^{\infty}_{0} \frac{d\omega}{\omega}\Delta\sigma'(\omega) \ .
\end{align}
The leading contribution to $\Delta\sigma'(\omega)$ comes from a tree-level process $\gamma \ell \to G \ell$, relevant diagrams are shown in Fig.~\ref{gravitonphotoprod}, from which one can deduce that $\Delta\sigma'_{\gamma \ell \to G \ell} \sim \cO(G_N\alpha)$. 
Therefore, using Eqs.(\ref{GDHint}) and (\ref{BG}), one arrives to a contradiction, since the LHS of Eq.(\ref{GDHint}) starts at order $\cO(\alpha G_N)$, while the RHS is only of order $\cO(G_N)$. This issue was first observed in Ref.~\cite{Goldberg:1999gc}, where the calculations of the cross section were done in the massless fermion limit. In the next section, we find $\Delta\sigma_{\gamma \ell \to G \ell} $ for the finite fermion mass.

\subsection{Asymmetry in Graviton Photoproduction: Spin-flip Cross Section}

The graviton photoproduction amplitude can be factorized as follows \cite{Holstein:2006bh}:
\begin{align}\label{factor}
&\langle p_f; k_f, \vec{\e}_f\vec{\e}_f|T|p_i;k_i,\vec{\e}_i \rangle = H \times \e^*_{f\mu}\e_{i\nu}T^{\mu\nu}_{\rm Compton} \ , \\[5pt] 
&H = \frac{\kappa}{4e}~\frac{(\e^*_f \cdot p_f) (k_f \cdot p_i) - (\e^*_f \cdot p_i) (k_f \cdot p_f)}{k_i \cdot k_f} \ ,
\end{align}
where $p_i$, $k_i$ are initial momenta of lepton and photon, $p_f$, $k_f$ are final momenta of lepton and graviton ($G$), $\e_{i \mu}(\lambda)$ is initial polarization of photon with helicity $\lambda$, $\e_{\mu\nu} = \e_{f \mu}\e_{f\nu}$ is the final polarization of the graviton written as a product of two `photon' polarizations.\footnote{We use conventions, where $\kappa^2 = 32\pi G_N = 8/M^2_P$. Additional conventions can be found in Appendix \ref{Example2}.} Finally, $T^{\mu\nu}_{\rm Compton}$ is the Compton scattering amplitude with final photon having polarization $\e_{f\mu}$. The relevant tree-level diagrams contributing to the graviton photoproduction are presented in Fig.~\ref{gravitonphotoprod}. It can be shown that, in the laboratory frame:\footnote{In the laboratory frame, we orient the $z$-axis along the direction of the incoming photon, and have $p_i = (m,\vec{0})$, $k_i=\omega_i(1,0,0,1)$, $k_f = \omega_f(1,\sin\theta,0,\cos\theta)$, $\e_i = (0,1,i\lambda,0)/\sqrt{2}$, $\e_f = (0,\cos\theta,\pm i,-\sin\theta)/\sqrt{2}$.}
\begin{align}
|H_{\rm lab}|^2 = \frac{\kappa^2 m^2}{8e^2} {\rm cot}^2\left(\theta/2\right) \ ,
\end{align}
where $\theta$ is the scattering angle of the graviton. 
%

\FIGURE{\includegraphics[width=11cm]{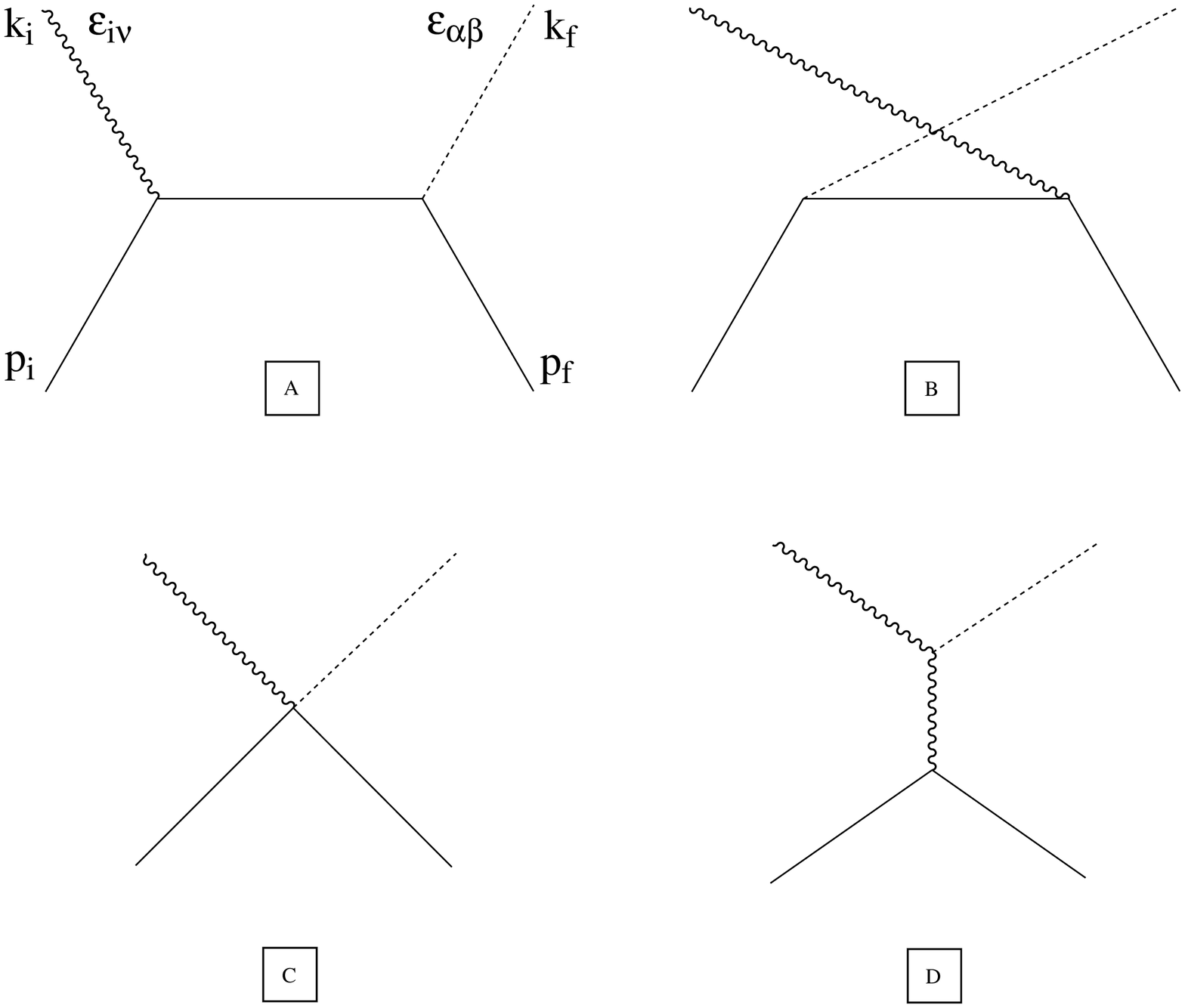}
 \caption{The tree-level diagrams contributing to the graviton photoproduction.}
\label{gravitonphotoprod}
}

According to Eq.~(\ref{factor}), to find $\Delta \sigma_{\gamma \ell \to G\ell}(\omega) $ in case of the graviton photoproduction, we multiply Eq.~(\ref{diffasymCompton})  by the additional factor $|H_{\rm lab}|^2$, and integrate over $d\Omega$. As a result:
\begin{align}\label{asym1}
\Delta \sigma_{\gamma \ell \to G\ell}(\omega) 
&= \frac{3\alpha \kappa^2}{16}\left(1+\frac{m}{3\omega}\right)\left[\frac{m}{\omega}\ln\left(1+\frac{2\omega}{m}\right) - 2\left(\frac{1+ \omega/m}{1+2\omega/m}\right)\right] \ .
\end{align} 
In the limit $\omega/m \gg 1$, we have $\Delta \sigma_{\gamma \ell \to G\ell}  \to -3\alpha/(2M^2_P)$, and in the limit, $\omega/m \ll 1$, we have $\Delta \sigma_{\gamma \ell \to G\ell} \to -2\alpha\omega/(3mM^2_P)$. In case $m \to 0$, we reproduce corresponding result of Ref.~\cite{Goldberg:1999gc} (notice, that we use different notations for $M_P$).

The GDH sum rule (\ref{GDHint}), can be rewritten as:
\begin{align}\label{modsumrule}
\frac{7\alpha}{16\pi^3}~\frac{m^2}{M^2_P}  +\frac{49}{256\pi^4}\frac{m^4}{M^4_{P}}  = \frac{m^2}{2\pi^2 \alpha}\left[\int^{\Lambda}_{0} \frac{d\omega}{\omega}\Delta\sigma'(\omega) + \int^{\infty}_{\Lambda} \frac{d\omega}{\omega}\Delta\sigma'(\omega) \right] \ ,
\end{align}
where $\Lambda$ is the UV cutoff above which the gravity becomes strongly coupled (in principle, $\Lambda$ can be much smaller than $M_P$). We will consider the first integral on the RHS, and will discuss the second integral in the next sections.

Substituting (\ref{asym1}) into the first integral in Eq.(\ref{modsumrule}),  we get:
\begin{align}\nonumber
I_{GDH} \equiv \frac{m^2}{2\pi^2 \alpha}\int^{\Lambda}_{0}\frac{d\omega}{\omega}~\Delta \sigma_{\gamma \ell \to G\ell}(\omega) &= \frac{m^2\kappa^2}{32\pi^2}\left[5 + \frac{1}{x}  - \frac{1+6x + 6x^2}{2x^2}\ln\left(1+ 2x\right)\right]  \\ \label{problem1} &= -\frac{3}{4\pi^2}\frac{m^2}{M^2_P}\left[\ln\left(\frac{2\Lambda}{m}\right) - \frac{5}{3} + \cO(x^{-1}\ln x)\right] \ ,
\end{align} 
where $x \equiv \Lambda/m \gg 1$. At this stage, it looks that we might have a problem, since $I_{GDH} \sim \cO(G_N)$, while the LHS of (\ref{modsumrule}) starts at order $\cO(\alpha G_N)$. Moreover, Eq.~(\ref{problem1})  contains a logarithmic term (divergent as $\Lambda \to \infty$).\footnote{It appears that even if we compute $\Delta \sigma_{\gamma \ell \to G\ell}$ in the massless limit, and assume $\omega > \omega_{th}$ in the $I_{GDH}$ integral, we would arrive to a same conclusion \cite{Goldberg:1999gc}, if we select $\omega_{th} \sim m$.}

We note in passing that $\alpha G_N$ corrections in (\ref{modsumrule}) come from: the tree level diagrams $\gamma \ell \to \gamma \gamma G \ell$, and from the interference of the tree-level amplitude $\gamma \ell \to G \ell$ of order $\cO(e \kappa)$, with its one-loop photon correction, of order $\cO(e^3\kappa)$, leading to order $\cO(\alpha^2 G_N)$ contribution to the cross section.

\section{One-loop Graviton Corrections to the Compton Scattering}

In this section we will show that there is no reason to doubt the validity of the GDH sum rule at the one-loop graviton level. The main reason leading to the problem was the assumption that one can find the forward Compton scattering amplitude simply by substituting the anomalous magnetic moment with the sum $a^{QED}+a^{QG}$. In fact, this is not true since there are other one-loop graviton diagrams that contribute to the spin-flip amplitude at order $\cO(G_N)$. The real parts of these loop diagrams give exactly the log-term (\ref{problem1}) obtained on the RHS of GDH sum rule. To show this result explicitly, we need to compute all one-loop corrections to the forward spin-flip scattering amplitude. An example, where analogous situation takes place is considered in Appendix \ref{Example}.


\FIGURE{\includegraphics[width=11cm]{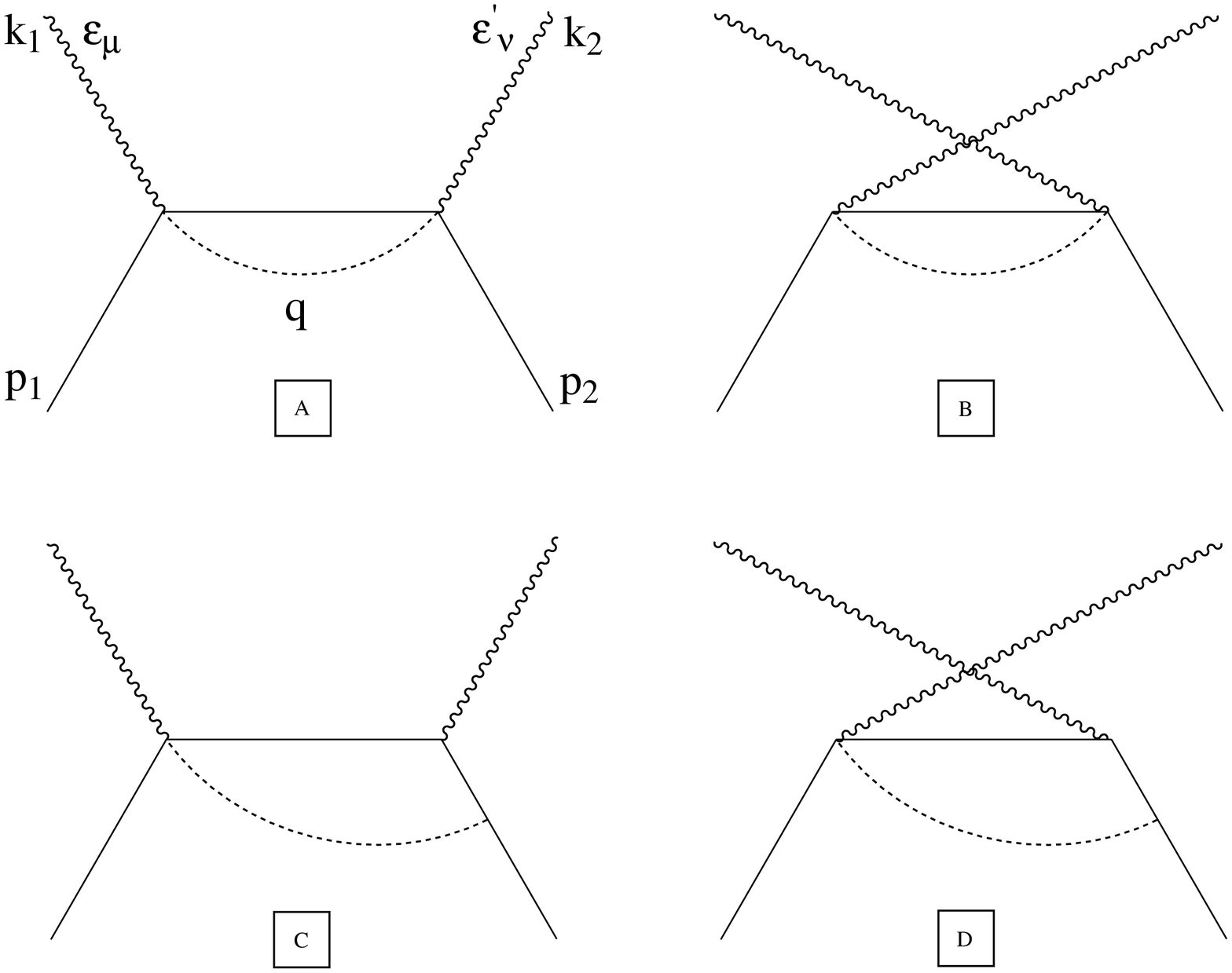}
\caption{One-loop diagrams that contribute to $f_-(\omega^2)$. There are other diagrams not shown here.}
\label{GravloopCompton}
}

The amplitude $f_{-}(\omega^2)$ receives not only contributions that result to the term proportional to $(a^{QED}+a^{QG})^2$, as was previously expected, but it also receives additional contribution from the other one-loop graviton diagrams. The examples of corresponding diagrams are given in Fig.~\ref{GravloopCompton}.  In particular, one can show that diagrams $A$ and $B$ in Fig.~\ref{GravloopCompton} give non vanishing contribution to the amplitude $f^{QG}_{-}(\omega^2)$, see Appendix \ref{Example2}. Using Pauli-Villars (PV) regularization scheme, from the computations analogous to the one presented in Appendix {\ref{Example2}, one can show that:\footnote{The reason we chose PV regularization scheme is that it is most suitable for comparison with the RHS of the GDH sum rule. The relation with dimensional regularization is straightforward.}
\begin{align}
f^{QG}_{-}(\omega^2 \to 0) = -\frac{3}{8\pi^2}\frac{\alpha}{M^2_P} \ln(\Lambda_{PV}/m) + \cO(\omega \alpha/M^2_P)  \ ,
\end{align}
where $\Lambda_{PV}$ is the PV regulator. Multiplying this amplitude with $2m^2/\alpha$, we will reproduce the logarithmic term of (\ref{problem1}}). In other words we simply verified the fact that the dispersive integral on the RHS of the GDH sum rule, reproduces the one-loop graviton corrections to the low energy forward scattering amplitude. To get an exact matching, we need: $\Lambda_{PV} = \bar{\Lambda} \equiv 2e^{-5/3}\Lambda$.\footnote{Note, that although the total one-loop graviton correction to the anomalous magnetic moment is finite, the same is not true for the Compton scattering amplitude.}
%


\section{Nonperturbative Regime}

Here, we will separate the second integral on the RHS of (\ref{modsumrule}) into two parts:
\begin{align}
\int^{\infty}_{\Lambda} \frac{d\omega}{\omega}\Delta\sigma'(\omega) = \int^{\omega_c}_{\Lambda} \frac{d\omega}{\omega}\Delta\sigma'(\omega) + \int^{\infty}_{\omega_c} \frac{d\omega}{\omega}\Delta\sigma'(\omega)   \ ,
\end{align}
where at energies $\omega > \omega_c$, for certain value of $\omega_c$, the black hole production would dominate, while for $\Lambda < \omega < \omega_c$ the string state exchange is expected to be important. We start by considering the second integral. Our discussions will carry semi-qualitative character.

\subsection{Black Hole Exchange}

Consider the possibility that as a result of $\gamma \ell$ scattering, the black hole state is produced with mass $M$, charge $Q$ and angular momentum $J$. In such case, the cross section can be crudely approximated by the cross sectional area of a black hole~\cite{gidd}:
\begin{align}
\sigma_J \sim \pi r^2_+ \approx 4\pi G^2_N M^2-2\pi G_NQ^2 - 2\pi \frac{J^2}{M^2}
\end{align}
where $r_+ = G_N M + \sqrt{G^2_NM^2 - J^2/M^2 -G_NQ^2}$ is the outer horizon of the Kerr-Newman black hole, and we assumed $J^2 + G_N M^2 Q^2 \ll G^2_N M^4 $.  Clearly, the above estimate of the cross section is only applicable in case $r_+ \sim G_N M \geq 1/\Lambda$, where $\Lambda$ is a cutoff of the theory (as was mentioned above). We will introduce a new parameter $\lambda_c$, so that: $\Lambda = \lambda_c M_P$, where $\lambda_c < 1$. Therefore, our estimate is valid, only when $M > \omega_c$, where $ \omega_c \equiv M_{P}\left(M_P/\Lambda\right) = \Lambda/\lambda^2_c $. In this case, the difference between the spin aligned ($J=j+1$) and anti-aligned ($J=j-1$) cross sections is:
\begin{align}
\Delta \sigma_{j}(\omega > \omega_c) \equiv \sigma_{j-1} -\sigma_{j+1} \approx \frac{8\pi j}{\omega^2} \ .
\end{align}
Therefore, the second piece of the  dispersion integral with $j=1/2$ will give:
\begin{align}
\frac{m^2}{2\pi^2 \alpha}\int^{\infty}_{\omega_{c}} \frac{d\omega}{\omega}\Delta\sigma_{\gamma \ell \to BH}(\omega) \approx \frac{1}{\pi}  \frac{\lambda^2_c}{\alpha} \frac{m^2}{M^2_P} 
\ .
\end{align}
As a result, 
\begin{align}\label{newsumruleB}
\frac{m^2}{2\pi^2 \alpha}\int^{\infty}_{0} \frac{d\omega}{\omega}\Delta\sigma'(\omega)  \approx -\frac{3}{4\pi^2}\frac{m^2}{M^2_P} \ln\left[\frac{\bar{\Lambda}}{m}\exp{\left(-\frac{4\pi \lambda_c^2}{3\alpha}\right)}\right] +\frac{m^2}{2\pi^2 \alpha}\int_{\lambda_cM_P}^{M_P/\lambda_c}\frac{d\omega}{\omega}~\Delta \sigma' \ .
\end{align}
In case, $\lambda_c \sim 1$, the integral on the RHS of (\ref{newsumruleB}) can be ignored, however, the first term becomes $\sim \ln(\bar{\Lambda}/\Lambda_L)$, where $m(\Lambda_L/m)^{9/8}$ is the QED Landau pole for one flavor. On the other hand, if $\lambda_c \sim \alpha$, the first term is $\sim \ln(\bar{\Lambda}/m)$,\footnote{Notice, that if $\lambda \sim \alpha$, then the BH contribution is of the same order as the first term in Eq.~(\ref{modsumrule}).}, and we can not ignore the integral on the RHS. Therefore, in this case, we need to consider the contribution from the remaining integral:
\begin{align}
I_S \equiv \frac{m^2}{2\pi^2 \alpha}\int^{\omega_c}_{\Lambda} \frac{d\omega}{\omega}\Delta\sigma'(\omega)  \ .
\end{align}

\subsection{String Regge Trajectory Exchange}

When $\lambda_c \ll 1$, we have a wide (energy) range where the gravity is nonperturbative, while the black hole formation is suppressed. 
In string theory, we might be in a regime, where $\lambda_c$ is associated with the string coupling constant, and $\lambda_c \ll 1$, therefore, $\Lambda = \lambda_cM_P$ should be associated with the string mass scale.\footnote{Type II superstring compactified on a 6-torus is an example of this general situation. In this case the relation between the string mass scale $M_S\sim 1/\sqrt{\alpha'}$ and the Planck scale is $M_P=M_S^4\sqrt{V_6}/g_S$, where $V_6$ is the volume of the 6-torus, and $g_S$ is the string coupling constant. In the perturbative regime, where $V_6M_S^6 \gg 1$ and $g_S\ll 1$, we have: $M_P\gg M_S $. When all radii of the torus are $\cO(1/M_S)$, $M_S$ becomes the only UV cutoff scale and $M_n=\cO(M_S)$.} 
In such case, we expect the range $\Lambda < \omega < \omega_c$ to be dominated by the string state exchange. 

Assume that for $\omega > \Lambda$, the 4-point string amplitude in the Regge limit has a non vanishing contribution to the spin-flip Compton amplitude. The fact that there might be such a contribution follows from the string Compton amplitude that was computed at tree and one-loop levels in a 4D fermionic heterotic string model \cite{Pasquinucci:1997gv}.
In general, the spin-flip cross section can be written as:
\begin{align}
\Delta\sigma_R(s) = \frac{4\pi r}{M^2_P}~(\alpha's)^{\alpha_R-1} \ ,
\end{align}
where $r$ is some (model dependent) constant, and we assumed for the scattering amplitude that $\Im \cA_R(s,t=0) \sim  \lambda^2_c~(\alpha's)^{\alpha_R}$. Here $\alpha_R$ and $\alpha'$ are Regge intercept and slope respectively, and we took into account that $1/M_P \sim \lambda_c\sqrt{\alpha'}$. In this case, if $\alpha_R =0 $, we will have:\footnote{
In general, 
%
$I_S \approx  \frac{r}{\pi \alpha} \frac{m^2}{M^2_P}~\frac{1 - \lambda_c^{4[1-\alpha_R]}}{1-\alpha_R} $ for $\alpha_R \neq 1$ and $I_S \approx  \frac{r}{\pi \alpha} \frac{m^2}{M^2_P}\ln(1/\lambda_c^4)$ for $\alpha_R= 1$.}
%
\begin{align}\label{stringstates}
I_S \approx   \frac{r}{\pi}\frac{\lambda^2_c}{\alpha} \frac{m^2}{\Lambda^2} =\frac{r}{\pi\alpha} \frac{m^2}{M^2_P}   \ .
\end{align}
For $\lambda_c \ll 1$, the intermediate string exchange regime will dominate over the black hole production, and the RHS of Eq.~(\ref{newsumruleB}) will be: $-\frac{3}{4\pi^2}\frac{m^2}{M^2_P} \ln\left[\bar{\Lambda}/\Lambda_r\right]$, where $\Lambda_r \equiv m\exp{\left(\frac{4\pi r}{3\alpha}\right)}$. It is interesting to note that: $\Lambda_r \approx \alpha M_P$, for $4\pi r \approx 1.01672$ and $m \approx 0.5~ {\rm MeV}$.\footnote{If the matter is confined on a D3-brane, one expects, $\lambda_c \sim \alpha$.}
This result might imply that for specific values of $r$, or specific string theories, the log-divergent term can be softened or completely removed. However, we are not able to provide more rigorous justification for this claim.

\section{Case with Extra Dimensions}

In the model \cite{ArkaniHamed:1998rs}  with $4+n$ extra dimensions, the one-loop bulk contributions to the anomalous magnetic moment of the muon were computed by Graesser\cite{Graesser:1999yg}. 
It appears that, the corrections from a single Kaluza-Klein (KK) graviton and KK radion are found to be each finite. The full one-loop bulk contribution is obtained by summing over all the KK states, and results in a model-independent correction to $a$, independent of $n$, and depending only on the scale of the strong gravity in $4+n$ dimensions, $M_D$.
Each KK mode $n$ of a bulk graviton contributes to $a$. Taking into account $(4\pi  G_N)^{-1} = M_{D}^{n+2}R^n$, the contribution from the tower of KK gravitons can be well approximated by the formula \cite{Graesser:1999yg}: 
\begin{align}
a^{QG} \approx \frac{c}{8\pi^2}\frac{2\pi^{n/2}}{n \Gamma[n/2]}~\frac{m^2}{M^2_{D}} \ , \ \ \ \ \ c = 5 - \frac{4}{3}\frac{n-1}{n+2} \ ,
\end{align}
where the first contribution to $c$ is from the tower of spin-2 KK states, and the second is from the tower of KK radion states. Note that this result scales as $M_{D}^2$ and is independent of $n$.

The contribution from the entire tower of KK gravitons to the spin-flip cross section was computed by Goldberg, and (in the limit where the target is massless) it is:
\begin{align}\label{goldberg}
\Delta\sigma_{\gamma \ell \to G \ell}(s) = \frac{2\pi^{n/2}}{n \Gamma[n/2]}R^n \int^{\sqrt{s}}_0 \Delta \sigma_{\gamma\ell \to G \ell}(s,m^2)m^{n-1}dm = \frac{1}{8}\frac{\alpha}{M^2_D}\left(\frac{s}{M^2_D}\right)^{n/2}A_n \ ,
\end{align}
where $A_n$ are some finite coefficients, for example: 
\begin{align}
A_2 = \pi(-25/4 + (17/3)\ln(2/\delta -1)) \ ,
\end{align}
where $\delta$ is a collinear cutoff, such that $-1+\delta \leq \cos \theta \leq 1 - \delta$. In the limit, $\delta \to 0$, we can approximate $A_2 \approx  \frac{17\pi}{3}\ln(2/\delta) $. In general, for any $n$, one can check that: $A_n \sim \ln(2/\delta)$.

Substituting Eq.~(\ref{goldberg}) into the sum rule integral, one obtains
\begin{align}
\frac{m^2}{4\pi^2 \alpha}\int^{\Lambda^2}_{0} \frac{ds}{s}\Delta\sigma_{\gamma \ell \to G \ell}(s)   =  \frac{1}{16\pi^2}\frac{m^2}{M^2_D}\left(\frac{A_n}{n}\right)\lambda_D^n \ ,
\end{align}
where $\lambda_D \equiv \Lambda/M_D$. Notice, that in deriving this result the limit $\Lambda/m \to \infty$ was taken from the beginning.
%
%
As in the 4D case, the value of this integral is expected to reproduce the real part of the one-loop graviton correction to the spin-flip low energy forward Compton scattering amplitude. Indeed, on dimensional grounds the one-loop integral should be $\Lambda^n$ divergent in $4+n$ dimensions. It may look like the collinear divergence, and the high-energy one are of different nature; the former is always diverges as $\ln(2/\delta)$ and the later as $\Lambda^n$, in case $n>0$. We will show that in 4D the collinear and UV divergences are related.

Assume that the 4D graviton scattering angle is such that: $-1 \leq \cos\theta \leq 1-\delta$, where $\delta \ll 1$, in which case:
\begin{align}\nonumber
&I_{GDH} = \frac{m^2}{2\pi^2 \alpha}\int^{\Lambda}_{0}\frac{d\omega}{\omega}\int^{2\pi}_0 d\phi \int^{1-\delta}_{-1} d\cos \theta~\frac{d}{d\Omega}\Delta\sigma_{\gamma \ell \to G\ell}(\omega,\theta) \\[7pt] \label{collinlog2} &= \frac{m^2}{4\pi^2M^2_P}\left[\frac{1-\delta/2}{1+\delta x}\left(5 + \frac{1}{x} + \frac{\delta(1+ 3(4-\delta)x)}{2}\right) - \frac{1+ 6x+6x^2}{2x^2}\ln\left(\frac{1+2x}{1+\delta x}\right)\right]  \ .
\end{align} 
If we take $\Lambda/m \to \infty$ limit first, we will get:
%
$I_{GDH} \approx  -\frac{3m^2}{4\pi^2M^2_P}\ln\left(2/\delta\right)$,
%
however, if we first take $\delta \to 0 $ limit, we obtain:
%
$I_{GDH} \approx  -\frac{3m^2}{4\pi^2M^2_P}\ln\left(2\Lambda/m\right)$.
%
For these limits to commute, we need $\delta = m/\Lambda$. 


\section{Summary}

In this paper, we showed that GDH sum rule is satisfied at one-loop graviton level. For this we started by computing the spin-flip cross section for graviton photoproduction on a spin-1/2 target of finite mass. Then, by substituting this result into the GDH sum rule we obtained the one-loop graviton correction to the low-energy forward Compton scattering spin-flip amplitude, also computed explicitly using the perturbation theory. 
The fact that the sum rule is satisfied, at least at one-loop level, suggests that the scattering amplitudes involved in the derivation of the sum rule maintain analyticity and unitarity at one-loop level. This is something that is expected from the renormalizable theory, however, for nonrenormalizable theory such as gravity coupled to matter, the situation might be less trivial. 
For example, it might happen that one needs to account for the higher dimensional terms in the effective action to restore the lack of unitarity in the original theory. With this result, we suggest that at one-loop level no additional terms are required to make the theory unitary. At the same time one should keep in mind that we concentrated specifically on processes such as Compton scattering and graviton photoproduction. It might happen that when considering a more exotic scattering process, the unitarity at one-loop might not hold.\footnote{Although, we were not able to find such an example, we do not know of any fundamental reason why this possibility should be excluded.} Needless to say, we also do not say anything about the unitarity or analyticity at multiple-loop level, and in the regime where gravity is strongly coupled.

At the semi-qualitative level, we also considered the case when gravity becomes strongly coupled, above the energy scale $\Lambda \equiv \lambda_c M_P$. In case, $\lambda_c \sim 1$, the GDH integral receives dominant contributions from the black hole exchange, and:
\begin{align}
\frac{m^2}{2\pi^2 \alpha}\int^{\infty}_{0} \frac{d\omega}{\omega}\Delta\sigma'(\omega)  \approx -\frac{3}{4\pi^2}\frac{m^2}{M^2_P} \ln\left[\Lambda/\Lambda_L\right]  \ , \ \ \ \ \Lambda_L \equiv m\exp{\left(\frac{4\pi \lambda_c^2}{3\alpha}\right)} \ .
\end{align}
On the other hand, in case, $\lambda_c \ll 1$, the string Regge trajectory exchange becomes the dominant one, and we have:
\begin{align}
\frac{m^2}{2\pi^2 \alpha}\int^{\infty}_{0} \frac{d\omega}{\omega}\Delta\sigma'(\omega)  \approx -\frac{3}{4\pi^2}\frac{m^2}{M^2_P} \ln\left[\Lambda/\Lambda_r\right] \ , \ \ \ \  \Lambda_r \equiv m\exp{\left(\frac{4\pi r}{3\alpha}\right)} \ .
\end{align}
We observe that if the matter resides on a D3 brane, so that $\lambda_c \sim \alpha$, then: $\Lambda_r \approx \Lambda$, for $m \approx 0.5~ {\rm MeV}$ and $4\pi r \approx 1$.\footnote{The dependence on the lepton mass is not essential for this argument.} We speculate that there might be a nonperturbative string like mechanism that cancels the UV divergences in the perturbative sector. 
One way to verify or falsify the above speculation is to consider the string one-loop corrections to the Compton scattering amplitude that was computed in the Ref.~\cite{Pasquinucci:1997gv}. 
In particular, we expect that the one-loop corrections to the Compton scattering amplitude would not produce UV log-divergencies, that are present in our case.

Finally, we briefly discussed the case with extra dimensions, and observed that in 4D the form of the divergence depends on the order in which the massless and collinear limits are taken. In particular, for these limits to commute, we need $\delta = m/\Lambda \to 0$. 


\acknowledgments

I thank Massimo Porrati for illuminating discussions. This work is supported by NSF grant PHY-0758032. 


\appendix

\renewcommand{\theequation}{A\arabic{equation}}
  \setcounter{equation}{0}

\section{Simple Working Example}\label{Example}

To demonstrate our point, it is instructive to consider the example of $2 \to 2$ scattering in $\lambda \phi^4$ theory. But before let us write the dispersion relation for the scattering amplitude $\cM(s)$ for which $\Re\cM(\infty) = {\rm const} \neq 0$.  The Cauchy integral over the closed contour, that encloses the cuts and consists of infinite half circles $\cC_{\infty}$, above and below the complex $s$-plane, can be written as:
\begin{align}\label{disprel}
\cM(s) &= \frac{1}{2\pi i}\int_{\cC_{\infty}}\frac{\cM(s')}{s'-s}ds' + \frac{1}{2\pi i}\int_{{\rm cuts}}\frac{\cM(s')}{s'-s}ds'  \\ \nonumber
&= \cM(\infty)  + \frac{1}{\pi}\int_{s_{th}}^{\infty}\frac{\Im \cM(s')}{s'-s}ds' \ .
\end{align}
This is known as dispersion relation with ``subtraction at infinity.'' Now, in case of the $\lambda \phi^4$ theory, the $2 \to 2$ scattering amplitude at order $\cO(\lambda^2)$ can be written as:
\begin{align}\label{amplitude}
&\cA(s,t) = -\lambda + \frac{\lambda^2}{32\pi^2}(V(s)+V(t)+V(u))\ , \\ \nonumber
&V(p^2) =  \int^1_0 dx \ln\left[\frac{x\Lambda^2_{PV}/m^2 +(1-x) - x(1-x)p^2/m^2 }{1 - x(1-x)p^2/m^2}\right] \ ,
\end{align}
where $s$, $t$, $u$ are Mandelstam variables ($s+t+u=4m^2$), and the logarithmic term is coming from the one-loop diagram which is regulated using Pauli-Villars regulator $\Lambda_{PV}$. In the forward scattering limit (when $t=0$) we have: 
\begin{align}\label{difference}
&\Re\cA(4m^2,0) \approx - \lambda + 3\frac{\lambda^2}{32\pi^2}\ln \frac{\Lambda^2_{PV}}{m^2} \ , \\ 
&\Re\cA(s\gg 4m^2,0) \approx -\lambda + 2\frac{\lambda^2}{32\pi^2}\ln \frac{\Lambda^2_{PV}}{m^2}  \\
&\Im\cA(s,0) \approx  \frac{\lambda^2}{32\pi}\sqrt{1-\frac{4m^2}{s}}  \ ,
\end{align}
where we used that for $t=0$, the complex $s$-plane has cuts for $|\Re s|>4m^2$. 
As should be expected from the optical theorem, $\Im \cA(s,0) = 2\sqrt{s(s - 4m^2)}~\sigma_{\rm tot}(s)$, and the leading contribution to the cross section is coming from the tree-level amplitude $\cA(s,t) =-\lambda$ for which $\sigma_{\rm tot} = \lambda^2/(32\pi s)$. 

The dispersion relation (\ref{disprel}) can be written as:
\begin{align}\label{compare}
&\Re[\cA(4m^2,0) - \cA(s \gg 4m^2,0)] = \frac{1}{\pi}\int_{4m^2}^{\Lambda^2}\frac{\Im \cA(s',0)}{s'-4m^2}ds' \\ \nonumber &= \frac{\lambda^2}{32\pi^2}\int_{4m^2}^{\Lambda^2}\frac{ds' }{\sqrt{s'(s'-4m^2)}} \approx \frac{\lambda^2}{32\pi^2}\ln \frac{\Lambda^2}{m^2} \ .
\end{align}
Comparing Eqs.(\ref{difference}) and (\ref{compare}) we verify that the dispersive integral indeed reproduces the loop correction to the forward scattering amplitude if we take $\Lambda_{PV} = \Lambda$. Clearly, this is what one should expect from analyticity and unitarity.


\renewcommand{\theequation}{B\arabic{equation}}
  \setcounter{equation}{0}

\section{Interaction Vertices and an Example of One-loop Computation}\label{Example2}

We will compute the contribution from diagrams A and B in Fig.\ref{GravloopCompton}. To obtain graviton vertex functions let's start with the combined action for interacting fermions and photons,
\begin{align}\label{matter}
S = \int d^4x \sqrt{-g}\left[\frac{i}{2}\psi^{\dagger}\gamma^0\gamma^{\alpha}D_{\alpha}\psi-\frac{i}{2}\left(D_{\alpha}\psi\right)^{\dagger}\gamma^0\gamma^{\alpha}\psi - m\psi^{\dagger}\gamma^0 \psi -\frac{1}{4}F^{\mu\nu}F_{\mu\nu}\right] \ ,
\end{align}
where $D_{\alpha} = e^{\mu}_{\alpha}\left(\partial_{\mu} + ieA_{\mu} - \frac{i}{4}\sigma^{\beta\gamma}e^{\nu}_{\beta}\partial_{\mu}e_{\gamma\nu}\right)$.\footnote{As usual, $\sigma^{\mu\nu} = i[\gamma^{\mu},\gamma^{\nu}]/2$, and the objects $e^{\mu}_{\alpha}$ are vierbein fields, such that $\eta^{ab}e^{\mu}_a e^{\nu}_b = g^{\mu\nu}$ and $g_{\mu\nu}e^{\mu}_ae^{\nu}_b = \eta_{ab}$.} The Lagrangian density describing the interaction with the graviton field at the lowest order is:
%
$\cL_{1} = -\frac{1}{2}\kappa h^{\mu\nu}T_{\mu\nu}$, 
%
where $g_{\mu\nu} = \eta_{\mu\nu} + \kappa h_{\mu\nu}$ with $\eta_{\mu\nu} = {\rm diag}\{1,-1,-1,-1\}$, and 
\begin{align}
T_{\mu\nu} = &\frac{i}{4}\bar{\psi}\left(\gamma_{\mu}\partial_{\nu}+ \gamma_{\nu}\partial_{\mu}\right)\psi -  \frac{i}{4}\left(\partial_{\nu}\bar{\psi}\gamma_{\mu}+ \partial_{\mu}\bar{\psi}\gamma_{\nu}\right)\psi + F_{\alpha\mu}F^{\alpha}_{\nu} + \frac{1}{4}\eta_{\mu\nu}F^{\alpha\beta}F_{\alpha\beta} \\ \nonumber &- \frac{1}{2}e~ \bar{\psi}\left(\gamma_{\mu}A_{\nu} + \gamma_{\nu}A_{\mu}\right)\psi  \ ,
\end{align}
is the symmetric and conserved current, which can be obtained from (\ref{matter}) along with the application of the equations of motion. 

In general, the off-shell $h\bar{\psi}\psi$-vertex function can be written as:
\begin{align}
V_{\mu\nu}(p,p') = \frac{1}{2}\left[\gamma_{\mu}P_{\nu} + \gamma_{\nu}P_{\mu}\right] - \eta_{\mu\nu}\left[\sh{P} - m\right] \ ,
\end{align}
where $P= (p+p')/2$. On the other hand, the $hFF$-vertex function is:
\begin{align}
W_{\mu\nu\alpha\beta}(k,k') &= 
-\eta_{\alpha\beta}k'_{\mu}k_{\nu} - \eta_{\mu\nu}(k_{\alpha}k'_{\beta}+k'_{\alpha}k_{\beta}) + k_{\nu}(\eta_{\alpha\mu}k'_{\beta}+\eta_{\beta\mu}k'_{\alpha}) \\ \nonumber &+ k'_{\mu}(\eta_{\alpha\nu}k_{\beta}+ \eta_{\beta\nu}k_{\alpha}) - k \cdot k'(\eta_{\alpha\mu}\eta_{\beta\nu}+ \eta_{\alpha\nu}\eta_{\beta\mu}-\eta_{\mu\nu}\eta_{\alpha\beta}) \ .
\end{align}
To compute diagrams A and B in Fig.\ref{GravloopCompton}, we need the following interaction Lagrangian:
\begin{align}
&\sqrt{-g}\cL^{h}_{\psi A} = -\frac{1}{2}e\kappa ~a_{\mu\nu\alpha\beta}(\bar{\psi}A^{\mu}\gamma^{\nu}\psi) h^{\alpha\beta} \ , \\
&a_{\mu\nu\alpha\beta} = \eta_{\mu\nu}\eta_{\alpha\beta} - \frac{1}{2}\left(\eta_{\mu\alpha}\eta_{\nu\beta} +\eta_{\mu\beta}\eta_{\nu\alpha}\right) \ .
\end{align}
%
%
%
%
%
It is straightforward to compute that the parts of the amplitudes that will contribute to the asymmetry in the forward scattering (after simplifications) can be written as:
\begin{align}
&i\cM_A = -\frac{3i}{4}e^2\kappa^2\bar{\psi}_{\sigma}(p_2)\left[\vec{\Sigma} J'(p_1,k_1)\right]\psi_{\sigma}(p_1)~(\vec{\e}^{'*} \times \vec{\e}) \ , \\ \nonumber
&i\cM_B = -\frac{3i}{4}e^2\kappa^2\bar{\psi}_{\sigma}(p_2)\left[\vec{\Sigma} J'(p_1,-k_2)\right]\psi_{\sigma}(p_1)~(\vec{\e}\times \vec{\e}^{'*} )\ , \\ \nonumber
&J'(p,k) \equiv \int \frac{d^4q}{(2\pi)^4}~\frac{1}{q^2}~\frac{\left(\sh{k} - \sh{q}\right)}{\left(p+ k - q\right)^2 -m^2}  \ ,
\end{align}
where $\vec{\Sigma} = {\rm diag}\{\vec{\sigma},\vec{\sigma}\}$, and we took into account that $\e_0 =\e'_0=0$. In the low-energy forward scattering limit, we choose $\vec{\e} = \vec{\e}' = (1,i\lambda,0)/\sqrt{2}$ and $\psi_{\sigma}(p) = \sqrt{m}\{\xi_{\sigma,} \xi_{\sigma}\}^T$. As a result, we have:
\begin{align}
&\cM \equiv \cM_A + \cM_B= \frac{3}{2}\lambda m e^2\kappa^2\sigma_3\left[I(p_1,k_1) -I(p_1,-k_2)\right]  \ , \\ \nonumber
&I(p,k) \equiv -i\int \frac{d^4q}{(2\pi)^4}~\frac{1}{q^2}~\frac{k^0 - q^0}{\left(p+ k - q\right)^2 -m^2}  
= \frac{k_0}{16\pi^2}\int^1_0 dx~\ln\left[\frac{\Lambda^2_{PV} - x(p+k)^2}{m^2 - x(p+k)^2}\right] \ .
\end{align}
Now, since $\Lambda_{PV} \gg m$, and $(p+k)^2 = m^2 \pm 2m\omega$, up to order $\cO(\omega)$ the integral is always real. 

Finally, taking into account that $f_{\rm scat}(\omega,\lambda) = \cM/(8\pi m)$ and Eq.~(\ref{minusamplitude}), the asymmetry in the forward Compton scattering amplitude (with $\sigma_3 = +1$) can be written as follows:
%
%
\begin{align}
\frac{m^2}{2\pi^2 \alpha} \cdot 4\pi^2 f_-(\omega^2\to 0) \approx \frac{3}{\pi^2}\frac{m^2}{M^2_P}\ln\left(\frac{\Lambda_{PV}}{m}\right)
\end{align}
Other one-loop graviton diagrams contributing to the spin-flip Compton amplitude can be computed using the knowledge of the interaction vertices given above.




\begin{thebibliography}{9}


\bibitem{Berends:1974tr}
  F.~A.~Berends and R.~Gastmans,
  ``Quantum Gravity And The Electron And Muon Anomalous Magnetic Moments,''
  Phys.\ Lett.\  B {\bf 55}, 311 (1975).

\bibitem{Graesser:1999yg}
  M.~L.~Graesser,
  ``Extra Dimensions and the Muon Anomalous Magnetic Moment,''
  Phys.\ Rev.\  D {\bf 61}, 074019 (2000)
  [arXiv:hep-ph/9902310].

\bibitem{Adams&al} 
  A.~Adams, N.~Arkani-Hamed, S.~Dubovsky, A.~Nicolis and R.~Rattazzi,
  ``Causality, analyticity and an IR obstruction to UV completion,''
  JHEP {\bf 0610}, 014 (2006)
  [hep-th/0602178].

\bibitem{Gerasimov:1965et}
  S.~B.~Gerasimov,
  ``A Sum rule for magnetic moments and the damping of the nucleon magnetic
  moment in nuclei,''
  Sov.\ J.\ Nucl.\ Phys.\  {\bf 2}, 430 (1966)
  [Yad.\ Fiz.\  {\bf 2}, 598 (1965)].
 %
  S.~D.~Drell and A.~C.~Hearn,
  ``Exact Sum Rule For Nucleon Magnetic Moments,''
  Phys.\ Rev.\ Lett.\  {\bf 16}, 908 (1966).
  
\bibitem{Goldberg:1999gc}
  H.~Goldberg,
  ``Gravitons and the Drell-Hearn-Gerasimov sum rule: Support for large  extra
  dimensions?,''
  Phys.\ Lett.\  B {\bf 472}, 280 (2000)
  [arXiv:hep-ph/9904318].

\bibitem{Low:1954kd}
  F.~E.~Low,
  ``Scattering Of Light Of Very Low Frequency By Systems Of Spin 1/2,''
  Phys.\ Rev.\  {\bf 96}, 1428 (1954);
%
  M.~Gell-Mann and M.~L.~Goldberger,
  ``Scattering Of Low-Energy Photons By Particles Of Spin 1/2,''
  Phys.\ Rev.\  {\bf 96}, 1433 (1954).

\bibitem{Weinberg}
 S.~Weinberg, 
 ``Lectures on Elementary Particles and Quantum Field Theory,'' 
    Volume 1, Brandeis University Summer Institute 1970 
    (S. Deser, M. Grisaru and H. Pendleton, editors, M.I.T. Press, Cambridge, 1970).  
    
\bibitem{Grigoryan:2012xv} 
  H.~R.~Grigoryan and M.~Porrati,
  ``New Sum Rules from Low Energy Compton Scattering on Arbitrary Spin Target,''
  arXiv:1204.1064 [hep-th].

\bibitem{Altarelli:1972nc}
  G.~Altarelli, N.~Cabibbo and L.~Maiani,
  ``The Drell-Hearn sum rule and the lepton magnetic moment in the Weinberg
  model of weak and electromagnetic interactions,''
  Phys.\ Lett.\  B {\bf 40} (1972) 415.

\bibitem{Dicus:2000cd} 
  D.~A.~Dicus and R.~Vega,
  ``The Drell-Hearn sum rule at order $\alpha^3$,''
  Phys.\ Lett.\ B {\bf 501}, 44 (2001)
  [hep-ph/0011212].

\bibitem{Brown:1952eu} 
  L.~M.~Brown and R.~P.~Feynman,
  ``Radiative corrections to Compton scattering,''
  Phys.\ Rev.\  {\bf 85}, 231 (1952).

\bibitem{PDG12}
 J.~Beringer \textit{et al.} (Particle Data Group), 
 Phys.\ Rev.\ D{\bf 86}, 010001 (2012). 

\bibitem{Brodsky:1968ea}
  S.~J.~Brodsky and S.~D.~Drell,
  ``The Anomalous Magnetic Moment And Limits On Fermion Substructure,''
  Phys.\ Rev.\  D {\bf 22}, 2236 (1980).
  
\bibitem{Holstein:2006bh}
  B.~R.~Holstein,
  ``Graviton physics,''
  Am.\ J.\ Phys.\  {\bf 74}, 1002 (2006)
  [arXiv:gr-qc/0607045];
%
 N.~A.~Voronov, 
 ``Gravitational Compton effect and photoproduction of gravitons by electrons,''		
 Soviet Physics JETP, Vol. 37, p.953;
 %
  S.~Y.~Choi, J.~S.~Shim and H.~S.~Song,
  ``Factorization and polarization in linearized gravity,''
  Phys.\ Rev.\  D {\bf 51}, 2751 (1995)
  [arXiv:hep-th/9411092].
  
\bibitem{gidd} 
  S.~B.~Giddings,
  ``The gravitational S-matrix: Erice lectures,''
  arXiv:1105.2036 [hep-th].
  
\bibitem{Lykken:1996fj} 
  J.~D.~Lykken,
  ``Weak scale superstrings,''
  Phys.\ Rev.\ D {\bf 54}, 3693 (1996)
  [hep-th/9603133].
  
\bibitem{Pasquinucci:1997gv} 
  A.~Pasquinucci and M.~Petrini,
  ``On the Compton scattering in string theory,''
  Phys.\ Lett.\ B {\bf 414}, 288 (1997)
  [hep-th/9708131].

\bibitem{ArkaniHamed:1998rs} 
  N.~Arkani-Hamed, S.~Dimopoulos and G.~R.~Dvali,
  ``The Hierarchy problem and new dimensions at a millimeter,''
  Phys.\ Lett.\ B {\bf 429}, 263 (1998)
  [hep-ph/9803315].
  
  
\end{thebibliography}
\end{document}